\documentclass[a4paper]{article}
\usepackage[dvips]{graphicx}

\newcommand{\be}{\begin{equation}}
\newcommand{\ee}{\end{equation}}
\newcommand{\bel}[1]{\begin{equation}\label{#1}}
\newcommand{\bea}{\begin{eqnarray}}
\newcommand{\eea}{\end{eqnarray}}
\newcommand{\beal}[1]{\begin{eqnarray}\label{#1}}
\newcommand{\nn}{\nonumber}
\newcommand{\nin}{\noindent}
\newcommand{\xp}{x_{\bot}}
\newcommand{\kp}{k_{\bot}}
\newcommand{\Rp}{R_{\bot}}

\def\d{\partial}
\def\xp{x_{\perp}}

\begin{document}
\title{\bf Parton picture of inelastic collisions \\
at transplanckian energies}

\author{ {\sc O.V. Kancheli}\thanks{e-mail: kancheli@heron.itep.ru} \\
     {\it  Institute of Theoretical and Experimental Physics, }  \\
     {\it  B. Cheremushinskaya 25, 117 259 Moscow, Russia. } }
\date{}
\maketitle

\begin{abstract}
We propose a parton model of inelastic collisions at
transplanckian energies $E \gg G^{-1/2}$~, using the gravitons,
whose transverse momenta are cut at the Planck scale, as partons.
For this purpose we represent the gravitational shock-wave
accompanying the fast particle in terms of such partons and take
into account the higher order multiperipheral-like
contributions.   We argue that the internal part of this shock
plane contains the ``black'' disk of radius $R(E) \sim E^{1/2}$
filled by such hard partons with the Planck density $\sim G$.
When two fast particles collide the hard graviton production
comes from the region of intersection of their black disks. The
corresponding value of inclusive cross-section at the given
rapidity and the impact parameter is proportional to the area of
this region. The final state with such hard gravitons is unstable
relative to the long range gravitational repulsion, and this
leads to the creation of the multiperipheral chain of black holes
at later stages of the reaction. We discuss various details of
this picture including higher order corrections and there
connection to a purely classical approach; we also  consider
briefly possible changes when additional hidden dimensions are
present.

\end{abstract}
\setcounter{footnote}{0}

\section{\bf Introduction}
\vspace{1pt}

The dynamics of particles interaction at transplanckian energies
is rather poorly understood up to now, even qualitatively. It is
usually believed that at such energies the dominant mechanism of
the interaction is a gravitational one, so that its contribution
highly overcomes that from other interactions. There is a lot of
papers where this problem is considered from a different
direction - here we mention only some of them \cite{eath1} -
\cite{hooft}~. ~The renewal of interest to this topics in last
years \cite{banks} - \cite{gidd2} is probably connected with the
discovery of various brane world like scenarios with TeV-scale
gravity \cite{add}. If this will be realized, then  gravitation
mediated effects can be found already at future accelerator
energies.

When two particles with high center-of-mass system (c.m.s.)
energy $\sqrt{s} \gg G^{-1/2} \equiv m_p$ and impact parameter $B
< \sqrt{s} ~m_p^{-2}$ come close one to another, then the high
energy (mass) $=~\sqrt{s}$~ is concentrated for a small time  in
the compact region. For this energy the `static' radius of the
horizon of a black hole (BH) with the mass $\sqrt{s}$ is much
larger than the size of the region, where the energy is
concentrated. Then, naively, one can expect that in the process
of the collision (or even before), an event horizon
``surrounding'' this region will form, and after this horizon will
gradually (at time $\sim \sqrt{s}/m_p^2$) transform to the
boundary of a real BH with the mass $\sim \sqrt{s}$. But, from
the other hand, the colliding particles  are highly
ultrarelativistic and it is unclear if it is enough time that
such a horizon can appear in a causal way.

To refine (or reject) these conclusions number of approaches was
developed. The gravitational field of a fast particle with energy
$E$ is concentrated in a shock-wave plane of thickness $\sim 1/E$
and is usually represented by the Aihelburg-Sexel (AS) metric
\cite{as}. The process of a classical collision of two such AS
waves is considered in the number of papers. In details it was
firstly done in \cite{eath1} where the transition of one AS wave
through another is considered in the geometrical optics
approximation. It follows that due to the curving and later
focusing of AS planes the curvature singularities can appear. They
signal the possible creation of a BH in process. It is also
mentioned in \cite{eath1},\cite{eath2} that a numerical solution
of the Einstein equation for this process also point to a
possible appearance of curvature singularities. The other (but in
principle the close) method \cite{gidd1} is based on a
construction of a maximally large trapped surface in the AS disks
collision process and perhaps confirms the same conclusions.

Unfortunately the transition of the one AS disk through another
is a very complicated nonlinear process with the unstable
behaviour, and it is unclear if one can believe in answers based
on the geometrical optics approximation. The other trouble of the
simple classical approach is that the curvature in essential
parts of AS-disks is too high ~($\gg m_p^2$) ~-~ and thus all
higher order corrections in curvature to the effective Einstein
Lagrangian can be of the same order, and this, like in the case
of the string interaction, can change the situation drastically -
in particular it can soften high curvature effects and freeze them
on the $\sim m_p$ scale. Then all the AS-disk focusing picture
\cite{eath1} can change for $B < \sqrt{s}/m_p^2$, because the
different degrees of freedom are relevant.

The different approach to transplanckian  collisions is the
perturbation theory. The diagram with a one graviton exchange
leads to the cross-section fast growing with the energy
\bel{per}
\sigma_{el}(B<B_0) ~=~ \int \limits_{k^2_{\perp min}}  \frac{d
\sigma_{el}}{d k^2_{\perp}} ~d k^2_{\perp} ~\sim~ G^2 s^2
k^{-2}_{\perp min} ~\sim~ B^2_0 (\frac{s}{m_p^2})^2 ~,
\ee
where $k^2_{\perp min} = B_0^{-1}$ is the infrared cutoff. The
expression (\ref{per}) hardly violates the unitarity for $s/m_p^2
\gg 1$~ because the allowed maximum of $~\sigma_{el}(B<B_0) \sim
B_0^2$.~ The eikonalization of the one graviton exchange cures it,
and leads \cite{hooft} to the elastic amplitude
\bel{gsmat}
A(s,b) ~=~1 - e^{i\delta}~~~ ,~~~~\delta = 8Gs ~\ln
\frac{B_0}{b}~.
\ee
This amplitude has a simple physical interpretation at the
lab.frame of the one of colliding particles, where the phase
$\delta(s,b)$ coincides with the time delay of the target
(``test'') particle when it penetrates through the AS shock front
at the impact parameter $b$.

Other high order corrections to (\ref{gsmat}) are also considered
in the literature - they include the multiperipheral diagrams
with gravitons and some generalizations, similar to that
formulated in the gluon (BFKL) case~ \cite{lip2,amati2}. But no
conclusions going far beyond that given by (\ref{per}) and
(\ref{gsmat}) is reached so far this way. At the same time it is
difficult to apply directly such perturbative methods to the
region of ``small'' impact parameters $b < m_p^{-1}
(s/m_p)^{1/4}$~,~ where, as one can expect, the most interesting
effects take place. One needs sertan indirect (bypass) way which
at the same time preserves the main aspects of the high order
perturbation theory, encoded in multiperipheral diagrams and
reggeons.

In this paper we speculate how the transplanckian particle
collision process can look out in the parton picture. One can hope
that by this approach one can  take into account the perturbative
graviton mechanisms and also include implicitly essential
nonperturbative and string effects. The parton picture represents
nicely all the main aspects of high energy hadronic interactions,
it also shows and ``explains'' the space-time picture of the
process. It can naturally take into account all the  regge
specifics of the interaction and also that one contained in the
dual models. Therefore one can expect that similar methods can be
helpful for gravity induced high energy interactions, partially
because there is no big difference between graviton and gluon
degrees of freedom  on the string scale. If partons are chosen
successfully, this can give a full qualitative picture of the
process and help to find a more accurate way to consider such a
phenomenon.

Our ``main partons here are the gravitons with the mean
virtuality $\sim m_p$. We suppose that gravitons with higher
virtualities are suppressed in a wave function of the fast
particle by some string-like mechanism
\footnote{The string-bit model \cite{thorn,Suss} is in some
respects similar to the hard parton constructions we used above,
but it starts directly from a fast string state}.
And partons with smaller virtualities interact much slowly and
lead only to corrections and to infrared effects.

\nin The outline of paper is as follows.

\nin {\bf Section 2}.~~  We consider some classical aspects of the
collision relevant to our consideration. We also discuss the
popular opinion that the transplanckian gravitational interaction
at not large impact parameters can be adequately described in
classical terms .

\nin {\bf Section 3}.~~ Here we consider how the wave function of
the fast particle can look in terms of such partons-gravitons. We
start from the Weizs\"{a}cker-Williams (WW) like partons,
decomposing the classical AS field and then try to refine the
model by the inclusion of graviton interactions and cascading.
The hard parton-graviton density is large in a transverse disk
with the radius $\sim m_p^{-1}\sqrt{E}$, and we suppose that
inside of such a disk the 2D parton density is saturated at the
Planck scale.

\nin {\bf Section 4}.~~ We consider the collision of two such
disks and study main inelastic processes. The resulting final
state  depends essentially on the impact parameter of the
collision and consists from layers  multiperipheraly distributed
in rapidity and filled with gravitons. In every such layer the
relative velocities of gravitons are locally thermalized,  with
relative energies $\sim m_p$. The process of the creation of this
final state takes a long time $\sim m_p^{-2}\sqrt{s}$, and the
order in which these layers are created is system dependent, like
in the usual multiperipheral picture.

\nin {\bf Section 5}.~~ Then we consider the future time evolution
of such a multiparticle state, created in a parton disks
collision. The classical instabilities, due to a long range
attraction between particles with not large relative velocities,
can take place and lead to the creation of the ``multiperipheral''
chain of BH from particles collapsing in every layer.

\nin {\bf Section 6}.~~ In conclusion we discuss some questions
related to this picture.

\section{\bf Some classical aspects of collision}

The gravitational field of a fast particle with the mass $m \ll
m_p = G^{-1/2}$ can be found by boosting of the static Newton
field. For a very high energy $E \gg m_p$ it gives approximately
the same as the boosted Schwarzschild field
\bel{busted}
g_{\mu\nu}-g_{\mu\nu}^{(0)} ~\simeq~  \frac{2 G
P_{\mu}P_{\nu}}{m} \frac{1}{\sqrt{\xp^2 +\gamma^2(z - \beta
t)^2}}~~,
\ee
where $g_{\mu\nu}^{(0)}$ is the Minkowski metric,~ $P_{\mu}$ -
fast particle 4-momenta, $\gamma=E/m$, $\beta=p/E$~.~ By going to
the limit $\beta \rightarrow  1$~ and making an additional
singular in $t,z$ coordinate transformation, one usually
represents this metric in the AS form
\bel{as}
   ds^2 ~=~ dx^+ dx^- ~+~ 4GE \ln\xp^2 \delta(x^-) (dx^-)^2
   -(d\xp)^2~~,
\ee
where ~$x^+ \equiv  v =t+z,~x^- \equiv u = t-z,~\xp$~ are the
light-cone coordinates. Another method to become the AS metric
which can be simply applied also to a distributed light-like
matter ($\sim$ partons), is based on the old observation that for
a general right-moved plane metric
\bel{peres}
   ds^2 ~=~ dx^+ dx^- ~+~ f(x^-,\xp)~ (dx^-)^2 -(d\xp)^2
\ee
Einstein equations reduce to a simple Poisson form
\bel{puas}
   R_{- -} ~=~ \d^2_{\bot} f(x^-,\xp) ~=~ G~T_{- -}(x^-,\xp)
\ee
For a bunch of massless particles with energies $\varepsilon_n$
and coordinates $(z_n,~ x_{n\bot})$,  moving along the z axis,
with the energy-momentum tensor
\bel{}
T_{- -}(u,\xp) ~=~ \sum_n \varepsilon_n
\delta^2(\xp-x_{n\bot})
  \delta(u - z_i)~.
\ee
it follows, due to linearity of (\ref{puas}), that
\bel{pack}
   f ~=~ 4GE \sum_n \big(\frac{\varepsilon_n}{E}\big)
     \delta(x^--z_n) \ln(\xp-x_{n\bot})^2~.
\ee
For $n=1$ this reduces to the AS metric (\ref{as}). For one
point-like particle the curvature fields are concentrated in a
plane shock-front $\delta(x^-)$, like the electric and the
magnetic fields for a fast Coulomb particle, where the same
structure ~$\delta(x^-) \ln\xp^2$~ enters the expression for
potentials
\footnote{The often mentioned analogy of the AS type metric with
shock waves is very superficial. In usual shock waves the energy
is concentrated not in the front discontinuity but in a long layer
of the gas moving behind the front.}.
For the metric of type (\ref{peres}) the only nonzero components
of the curvature tensor are
\bel{curv} R_{-\bot -\bot} ~=~ -\d_{\bot} \d_{\bot} f(x_-,\xp)
~\sim~ \delta(x^-) \xp\xp/\xp^4~, \ee
where the last expression is for a point-like particle. The
curvature tensor for (\ref{busted}) has approximately the same
structure as (\ref{curv}), but is evidently smeared in $x^-$ on
$m\xp/E$ and, what is in a certain sense essential, its Weyl part
belongs to a different canonical type.

In connection with the previous is also such a property of the AS
metric (\ref{as}), and also of the more general one
(\ref{peres}), that the corresponding symmetrical energy-momentum
pseudotensor ~$\theta^{(LL)}_{\mu\nu}$ vanishes - so that it is
hard to define in a consistent way the energy contained in AS
type field
\footnote{This takes place not only for a Landau-Lifshitz form of
$\theta^{(LL)}_{\mu\nu}$. The other popular forms of
$\theta_{\mu\nu}$, for example the canonical one, also give zero.
This zero is the result of the change of the canonical Weyl tensor
type in the AS case ~-~ which by itself follows after the singular
coordinate transformation made while going from (\ref{busted}) to
(\ref{as}).}.
But for the boosted Schwarzschild metric (\ref{busted})  the
$\theta^{(LL)}_{\mu\nu}$ is nonzero ~-~ it has only one large
component
\bel{lali} \theta^{(LL)}_{--} ~\sim~ \frac{ G \gamma^2
m^2}{(\xp^2 +( \gamma x^-)^2)^2}~, \ee
as is normal for an ultrarelativistic object. The energy flow in
this case is
$$
\varepsilon(\xp) \simeq \int\limits^{\infty}_{\infty} dx^-
 \theta^{(LL)}_{--} ~\simeq~  \frac{1}{\xp^3}
\Big(\frac{m}{m_p}\Big) \Big(\frac{E}{m_p} \Big)~,
$$
and the quantity
$$
\varepsilon ~\simeq~ \int\limits^{\infty}_{1/m} d^2\xp
\varepsilon(\xp) ~\sim~  \Big( \frac{m}{m_p}\Big)^2 ~E
$$
gives the energy in the coherent gravitational field of fast
particles.

The massless test particles colliding with the AS disk on the
distance $\xp$ from the center will be captured (trapped) by a
disk for time ~$= 8GE\ln (L/\xp)$,~ and during this time
transported backward in $z$ to the distance $8GE\ln L/\xp$ and
then released from opposite side of the AS disk (See Fig.~1).
\begin{figure}[h]    \centering
\vspace{15mm}
\begin{picture}(70,110)(50,0)
\includegraphics*[scale=.3]{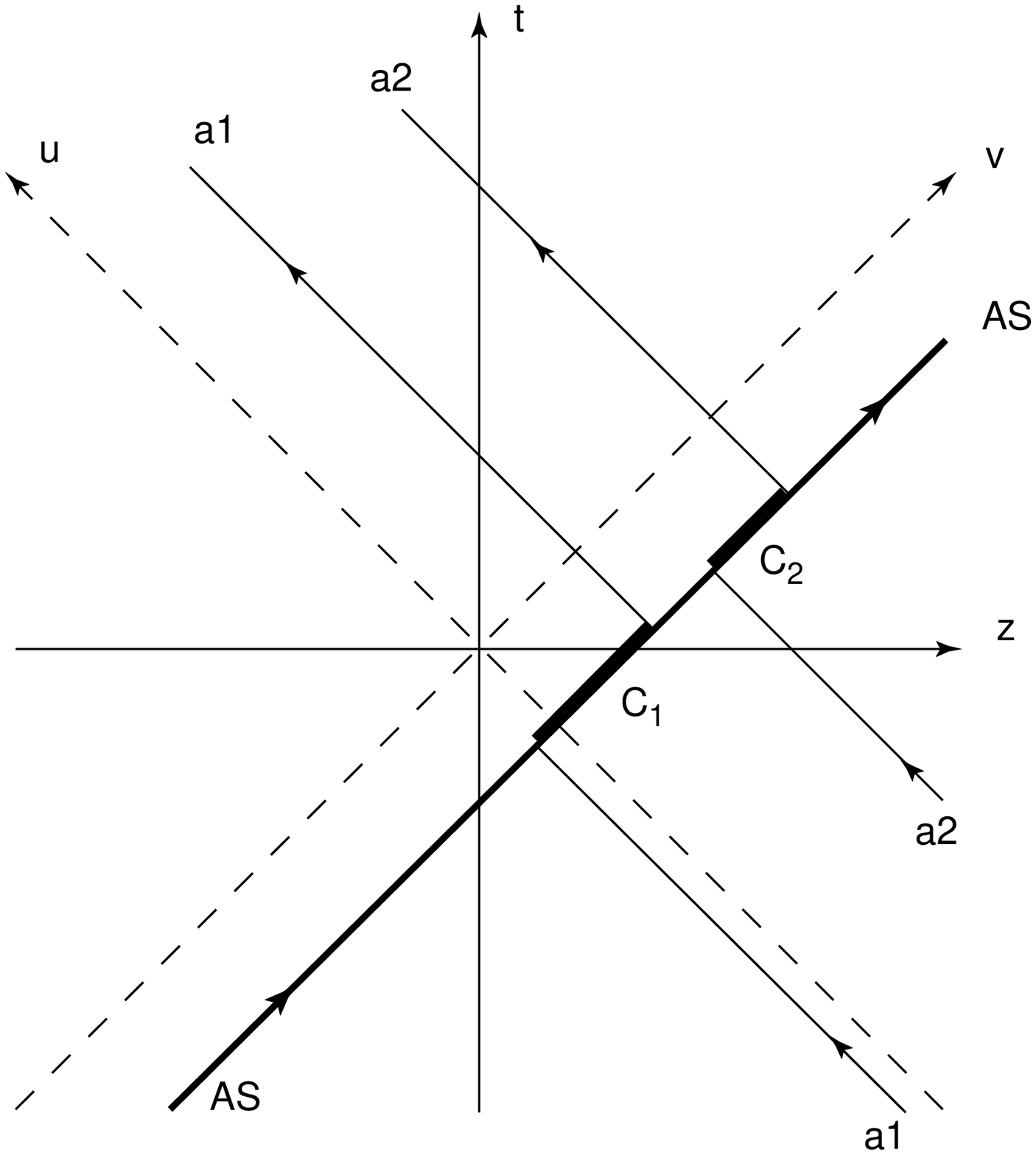}
\end{picture}
\parbox{10cm}{{\bf Fig.1~:}~~ Trajectories of ``test'' particles
$a_1$ and $a_2$ colliding with the AS disk accompanying a fast
transplanckian particle. On the sections $C_1$ and $C_2$ the
particles are ``transported'' by the AS disk.}
\end{figure}
After that particles continue there motion with the same momenta
as before the collision~
\footnote{Such a behavior is a simple reflection of that phenomena
that in the gravitational field  of a ``standing'' particle time
is delayed. So moving in this field massless particles are
slightly slow down (remain behind in $z$) near the mass on time
$\delta t \simeq 4Gm\ln (L/\xp)$, where $L\gg \xp$ is the full
distance in $z$ that particles traveled. As a result they remain
behind in z from the similar trajectories with larger $\xp$ If we
boost this time delay by gamma-factor $\sim E/m$ we come to the
space-time delay $4GE\ln (L/\xp)$, and this  corresponds to the
picture of the motion of the test particle captured by the AS
front.}.

Trajectories of these test particles look discontinuous (if we
``forget about sections $C_1$ and $C_2$ on Fig.1, where particles
simply move backward glued to the AS plane). One can perform the
specific coordinate transformation \cite{eath1}, singular in the
AS plane, so to make the test particles trajectories, crossing
the AS plane, continuous. Making it one cuts out a region of the
space-time around the AS plane (of $1/E$ thickness), after that
shifting and gluing together the distant planes. But this
procedure looks very strange and risky for a nonsingular metric
(\ref{busted}), for which the test particle trajectories are
continuous.

The classical collision of two AS disks is considered in the
number of works but no exact solution showing how they move one
through another was found. The most detailed analysis of this
process was first given in the paper of D'Eeath \cite{eath1} (See
also \cite{eath2} and papers cited there). One can mention three
main approaches to this problem. One is based on a numerical
solution of Einstein equation, another on the geometrical optics
approximation, and the third on the analysis of the possible
appearance of trapped surfaces during the process of the two AS
disks collision.

In the geometrical optics approximation one of the AS surface is
treated as a system of many infinitesimal test particles
(directrix) which cross the another AD disk and do not disturb
them during the traverse period. The same is symmetrically  done
with another AS-disk. Because the time-delay depends on $\xp$,
during such a transition of the AS fronts one through another,
they curve. At later times these curved AS disks focus on caustic
surfaces, where their curvature becomes very big, and this signals
about the creation of some BH like objects, possibly surrounded
by horizons. As mentioned in \cite{eath2}, the consideration of
the same problem by means of a numerical solution of Einstein
equations gives the reminding picture.

Another approach is based on a search of possible  trapped
surfaces shortly before the AS-disk's collide
\cite{eath1,gidd1,gidd2}. If such a surfaces exist, one usually
predicts that a black holes will be created in process of the
collision and even estimates their mass and the corresponding
cross-section for the inelastic interaction. Such a consideration
is usually done in specially transformed coordinates, singular
with respect to the ``normal'' one and chosen in such a way to
make the test particles trajectories ``continuous'', when they
cross the AS surface. This is a risky procedure because some
effects can arise due to the singular structure of AS planes and
the real answer can depend on that what takes place when two
singular AS planes go one through another in the period of their
collision, where we in fact do not know the metric.

At the same time one can try to consider this  in ``usual''
nontransformed coordinates. The picture is illustrated by Fig.2~
where the configuration of two disks is shown for some time
before the collision.
\begin{figure}[h]    \centering
\begin{picture}(70,130)(50,0)
\includegraphics*[scale=.3]{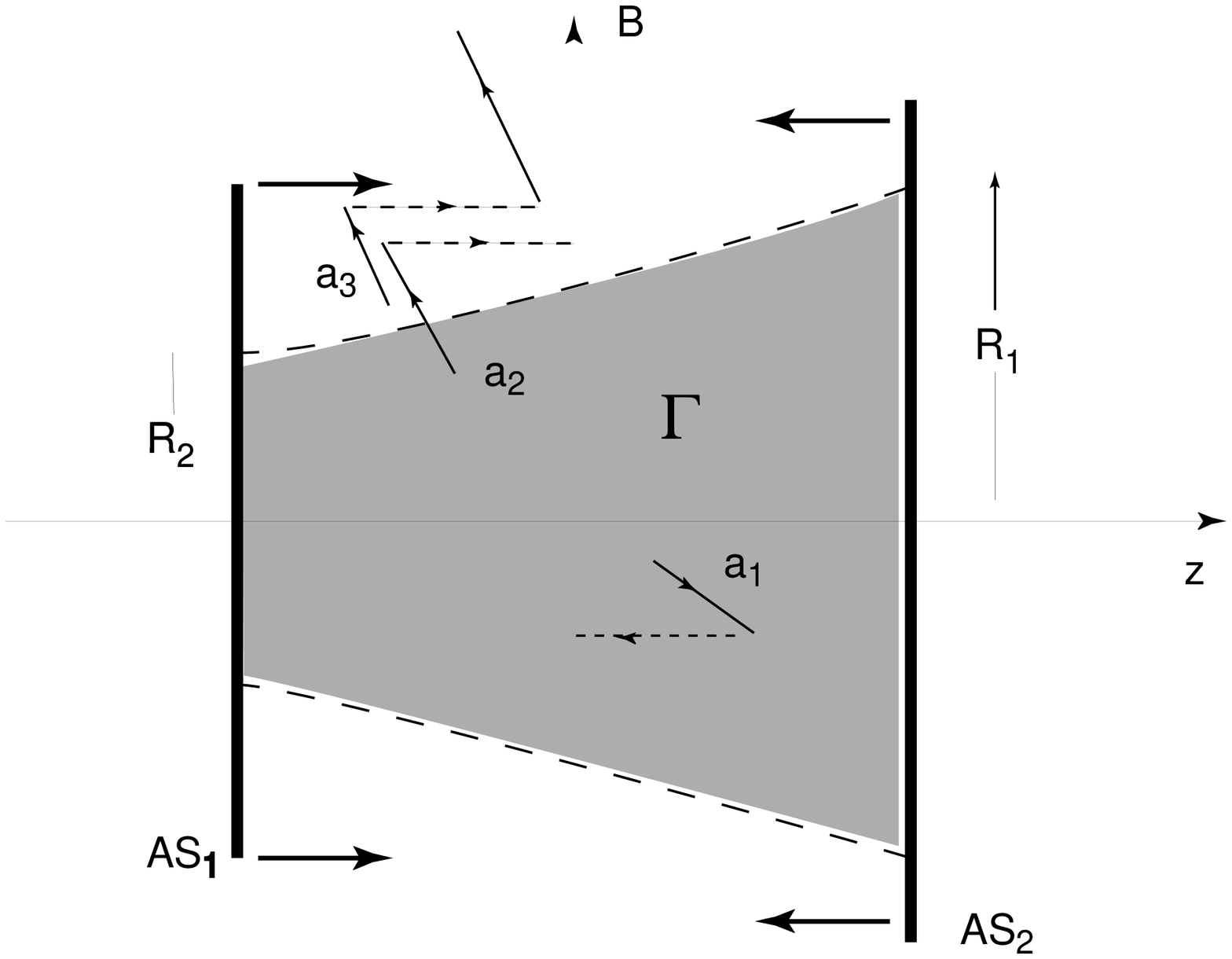}
\end{picture}
\parbox{10cm}{{\bf Fig.2~:}~~Particles from $\Gamma$ - the filled
region between approaching one to another AS disks can not escape
to infinity. They collide with disks $AS_1$ or $AS_2$ and after
that are transported to the plane $z=0$ where two disks
coincide(merge), before particle come out from other side of
disks. Particles $a_1$ and $a_2$ shown on Fig. are trapped and
particle $a_3$ have time to cross the $AS_1$ disk and escape. This
type condition in fact fully defines the shape of region $\Gamma$
for all $B$ and $y$.}
\end{figure}
The region $\Gamma$ is defined in such a way that the massless
test particle can not escape from this region before the AS disks
coincide. The space between disks is flat. But when a particle
will try to escape from the region $\Gamma$ it will collide with
one of the disks and will be absorbed (glued) and transported
with it towards another AS disk. The longitudinal and the
transverse size of $\Gamma$ can be defined from the condition that
the particle glued to one of $AS_i$ disks will not be released
(from another side of the disk) before disks collide. The region
$\Gamma$ is maximal in c.m.s. and for $B=0$,~ when it is cylinder
with the radius $\xp$ and the height $Z=2\xp = 8GE$. In the frame,
where particles have energies $E_1,~E_2$ but again $B=0$, the
region $\Gamma$ is a truncated cone with foundations $R_i = 4GE_i$
and $Z = 4G(E_1
 + E_2)$. For $B\neq 0$ the shape of $\Gamma$ is deformed (see
\cite{gidd2}).

But how can one conclude only from the existence of a region
$\Gamma$ with such a properties that anything singular happens in
the future evolution of the AS disk system? Evidently that this
future depends only on that what takes place on a complicated
nonlinear stage when two AS disks  (smeared in $z$ by $\sim
1/E_i$) coincide. And the existence of the empty region $\Gamma$
nothing changes in this respect. When regarded from such a
position, the trapped surface found in coordinates with
continuous trajectories can be the artifact of using the singular
coordinates. Another risky moment of using the trapped surfaces
for the prediction of the future system evolution is that these
trapped surfaces are isolated from the ``rest world'' by the the
AS shock fronts where the curvature is singular (or partially
coincide with the $AS_i$ planes). And the last moment. If for
colliding particles we use the regular metric (\ref{busted}),
instead of the AS, then the previous consideration of $\Gamma$
(like in Fig.2) remains unchanged. But if some trapped surfaces
exist in this case is rather unclear now and needs an additional
investigation. It is possible that the classical ``singular''
methods as used in \cite{eath1,eath2,gidd2} lead to more or less
adequate answer for the BH production, but the physics of this is
not evident.

\nin At the end of this section two additional remarks

\begin{itemize}
\item
The whole picture of the AS disks collision should be
longitudinally boost covariant, when we change the energies of
colliding particles $E_1 ~\rightarrow~ E_1 \xi$, ~$E_2
~\rightarrow~ E_2 / \xi$~ leaving $s \simeq 2 E_1 E_2$~ and the
impact parameter $B$ invariant. By this transformation we can
always select a frame in such a way that it is close to the
``laboratory'' frame, when one of the energies is very big $E_1
\gg m_p$~ and the another is ``arbitrary'' small $E_2 \ll m_p$.
Moreover, because the mass of the particles is ``in our hands'' we
can make the energy $E_2$ so small that it can be considered as a
``test particle'' with the respect to the particle $E_1$. In such
a system the ``trapping'' region $\Gamma$ disappears and all the
space-time before the $AS_1$ disk corresponds to the region of the
intersection of two disks in c.m.s. ($z=0$ in Fig.2). Now we have
only one AS disk and an additional light-like test particles can
move to infinity without any intersection with the $AS_1$ plane.
All this means that if we hope that some classical curvature
singularities must necessary appear in such a collision (remember
that $s \gg m_p^2$), then, in this system, they must originate
from an instability of the individual AS disk with the respect to
a small disturbances.

\item
The curvature in the AS shock front is high - for the
transplanckian case it is much larger than the critical one - for
some components of the curvature tensor
$|R_{\mu\nu\lambda\sigma}| \gg m_p^2$. Therefore in the effective
gravitational Lagrangian  the higher order terms in
$R_{\mu\nu\lambda\sigma}$ can be of the same order of the
magnitude (or even bigger) than the standard  $R$ term. Such
terms can be induced by the quantum fluctuations or ``come from
strings''. For the metric of type (\ref{peres}) this will
correspond to additional terms at the left hand side of
Eq.(\ref{puas}):
\bel{nlin}
 \d^2_{\bot} f(x^-,\xp)  ~+~
 c_2 m_p^{-2} \big(  \d^2_{\bot} f(x^-,\xp) \big)^2
  ~~+~~...~~    ~=~ G~T_{- -}
\ee
These terms can radically change the structure of AS solution for
$E \gg m_p$~, smoothing  the entering (\ref{pack}) singularities
\footnote{For one light particle state one can always, by moving
to a rest frame, make the contribution of additional terms in
(\ref{nlin}) small. But for colliding transplanck particles these
terms can be very essential, especially in a AS-disk's
intersection region.}.
Also, depending from signs and relative values of nonlinear terms
in (\ref{nlin}), this can lead, for example, to a saturation of
values of $f$ in the internal parts of AS disk, where $f$ in the
linear case can be arbitrary big.

\end{itemize}

\section{\bf  Parton structure of fast particle with $E \gg m_p$ }

To understand the AS shock fronts interaction, when the classical
picture is unreliable, we introduce the ``microscopic'' model of
the AS disk in terms of parton degrees of freedom. There is a
number of possibilities for a choice of constituents (partons)
Various degrees of freedom or there combinations can be used as
partons but it is very difficult to choice between them the most
adequate one. It is possible the string coordinates (or string
bit's \cite{thorn}) are the most appropriate ones. But the
dynamics at the high string density is also unclear - it is
possible here again we come to a particle-like media. We use the
transverse gravitons as partons, and regulate their spectra at
$\kp \geq m_p$ in a way suggested by ``strings''. Going to Fourier
components of the metric
\bel{four} \sum_{\lambda}
a^{\lambda}(k)\epsilon^{\lambda}_{\mu\nu} ~=~ \omega \int ~d^3x
~e^{ikx}~  (~ g_{\mu\nu}(x)- g_{\mu\nu}^{(0)}(x)~)~,
\ee
where $\epsilon^{(\lambda)}_{\mu\nu}$ -gravitons polarization
tensors, ~$\omega=\sqrt{k_z^2+\kp^2}$~, and substituting into the
right hand side of (\ref{four}) the AS metric components from
(\ref{as}) or from (\ref{busted}) we find:
\bel{lt} \sum_{\lambda} a^{\lambda}(k)\epsilon^{(\lambda)}_{--}
~=~ ~\frac{E}{m_p} ~\omega~\frac{1}{\kp^2}~. \ee
From the transversality condition
~$k^{\mu}\epsilon^{(\lambda)}_{\mu\nu} = 0$~ in the gauge
~$\epsilon^{(\lambda)}_{+\mu}~=0$~ we have the following relation
between longitudinal and transverse projections of polarization
tensor~:
\bel{tra}
  \epsilon^{(\lambda)}_{--} ~=~ \epsilon^{(\lambda)}_{\bot\bot}
  \frac{(k^+)^2}{\kp^2}~.
\ee
With the fields normalization used in (\ref{four}), the full
parton (graviton) number in a fast particle field is given by :
\bel{pn} N~=~ \int d^3k~ n(k) ~=~ \int \frac{d^3k}{\omega}
\sum_{\lambda}~a^{\lambda}(k) a^{+\lambda}(k)~.
\ee
Then, substituting (\ref{lt}) and (\ref{tra}) in (\ref{pn}), we
can extract the transverse graviton density in the AS disk :
\beal{grsp}
dn^{\bot}(E, \omega,\kp) ~\sim~ \frac{(a^{\lambda})^2}{\omega}~
  d\omega d^2\kp ~\sim~~~~~~~~~~~~~~~~~~~~~~~~~~~~~  \nn  \\
 ~~~~~~~~~\sim~ \Big( \frac{E}{m_p} \Big)^2
  ~\frac{\omega}{(k^+)^4}~ dk_z d \kp^2
  ~\sim~ \Big( \frac{E}{m_p} \Big)^2
  ~\frac{d\omega}{\omega^3}~d^2\kp ~~,
\eea
These spectra differs crucially from the QED vector parton spectra
\footnote{The parton spectra (\ref{grsp}) can be found also from
the WW like factorization of one graviton exchange diagram as is
usually done for photons in QED books.~ It is instructive to
compare the expression (\ref{grsp}) with the general form of WW
spectra for the spin J field
$$
d n^{\bot} ~\sim~ g_J^2~ \Big(\frac{k_{\bot}E}{\omega} \Big)^{2J}
\Big( \frac{\omega d\omega}{E^2} \Big) ~\frac{d^2
k_{\bot}}{k_{\bot}^4}~~=~
g_J^2~ \frac{dx}{x^{2J-1}}~\frac{d^2\kp}{\kp^{2(2-J)}}~~,
$$
where $g_J$ is the corresponding coupling constant :~~~$g_1
\leftrightarrow e_{qed},~ g_2 \leftrightarrow \sqrt{G}~$,...~and
$x$ - Feynman scaling fraction $\omega/E$. },~
by the $\omega$ and $\kp$ dependence and by the overall growth of
the parton density $\sim E^2$.

In a ``full'' theory one can expect that the spectra (\ref{grsp})
are essentially suppressed at $\kp ~>m_p$ due to various nonlocal
`string' effects. Here we simply suppose that they are cut at
$\kp \geq \kappa \sim m_p$. Than the following integrated over
$\kp$ spectra
\bel{spsp}
dn^{\bot}(\omega) ~=~  d \omega \cdot
\int\limits_0^{\min(\omega,\kappa)} d^2\kp n^{\bot}(\omega,\kp)
~\sim~ \Big(\frac{E}{m_p}\Big)^2
\frac{\kappa^2}{\omega(\omega^2+\kappa^2)}~d\omega
\ee
can be approximately used for all $\omega$.

Now let us discuss how essential are the higher order corrections
to these primary parton spectra. The main higher order diagrams at
high energies, which ``follow'' the WW contribution - are the
multiperipheral ones. In terms of partons they correspond to a
parton cascading which increases with $E$ the full number of low
energy partons, and as a result the cross-sections
\footnote{ The parton cascading gives the shift of the reggeon
intercept $\alpha(0)$ from their bare values. This corresponds to
an additional growth of the parton density by the factor $\sim
(E/\omega)^{\Delta}$, when we go to smaller parton energies
$\omega$. The average interval in rapidity between the parton
splitting (ladder rungs) is $\sim 1/\Delta$. In the BFKL
\cite{BFKL} case (vector partons) the shift of the pomeron
intercept $\alpha(0)$ from their bare value is $\Delta \sim
g_{QCD}^2$,~ but the gluon intercept probably does not shifts
because of the gauge nature of gluon. The gravitons intercept
probably also do not change, by the same reason.  But the
intercept of the bare 2G trajectory ( $~= 3$) corresponding to
(\ref{grsp}) can shift, if nothing prevents.}.
Such graviton ladder corrections for a scattering amplitude are
considered in a number of papers \cite{lip1} in the pure gravity,
supergravity and also for ``multiperipheral'' string amplitudes
\cite{amati2}.
\begin{figure}[h]    \centering
\begin{picture}(80,115)(130,0)    
\includegraphics*[scale=.4]{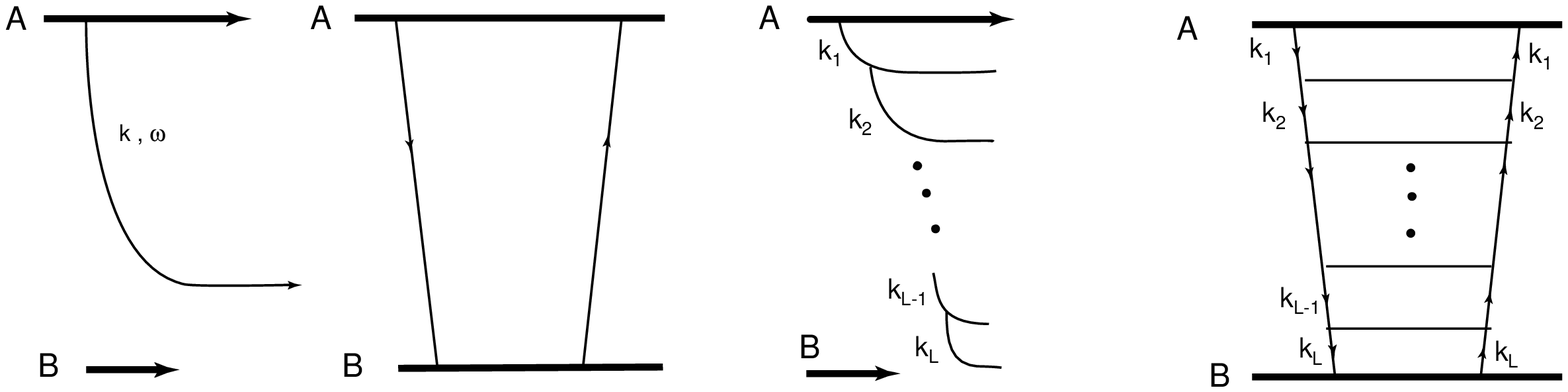}
\end{picture}
\parbox{10cm}{{\bf Fig.3~:}~~
The fast particle A with its parton cloud interacts with the
target~B~; (a) and (b) represent the WW amplitude and one cell
ladder corresponding to a WW~ cross-section~;~~ diagrams  (c) and
(d) represent the graviton cascading and the ladder amplitude
corresponding to a cross-section of the interaction with the
target}
\end{figure}
We will not review here these calculation - for our qualitative
consideration we only mention  some aspects essential for us.
Firstly we suppose that such ladders are supplemented by some
vertex factors $\Gamma(k_{i\bot}/m_p^2)$ that smoothly cuts high
virtual transverse momenta $k_{i\bot} > m_p$ on exchanged lines
(see Fig.3). Then for integrations over $k_{i\bot}$ in all cells
of the ladder the mean contribution comes from the region near
this upper bound $k_{i\bot} \sim m_p$.

The averaged picture of the ``ladder''-cascading process in parton
terms can be represented as a chain of $L$ step convolutions of
the ``primary'' WW spectra  $n^{\bot}(\omega_i,
\omega_{i+1},\kp)$~ given by (\ref{grsp}). This gives:
\beal{convol}
\int d\omega_1 d^2 k_{1} ~n^{\bot}(E,\omega_1, k_1)
\int d\omega_2 d^2 k_{2} ~n^{\bot}(\omega_1,\omega_2, k_2)~ ...
\nn \\
...~
\int d\omega_{L-1} d^2 k_{L-1} ~n^{\bot}(\omega_{L-1},\omega_L,
k_L) ~\sim \\
\sim~~ \frac{\Delta^L}{(L-1)!} ~\Big( \frac{E}{m_p} \Big)^2
~\frac{1}{\omega_L^3} \ln^{L-1}\Big(
\frac{E}{\omega_L}\Big)~~,~~~~~~~~~~~~~~~   \nn
\eea
where
$$
\Delta ~=~ m_p^{-2} \int d^2\kp \Gamma^2
\Big(\frac{\kp^2}{m_p^2}\Big)~,
$$
and where, as have mentioned above, we included the effective
graviton emission vertexes $\Gamma$ in the density $n^{\bot}$ to
take into account nonlocal effects  responsible for the cutoff of
high $\kp$. Summing over the number of cascade steps $L$ we come
to the ``reggeized'' graviton-parton spectra
\bel{grsp2}
dn^{\bot}(E, \omega,\kp) ~=~
 ~\Big( \frac{E}{m_p} \Big)^2 \Big(\frac{E}{\omega} \Big)^{\Delta}
 ~\frac{d\omega}{\omega^3}~d^2\kp ~~.
\ee
Such corrected parton spectra are even more concentrated
\footnote{The value of $\Delta$  depends strongly on the behaviour
of the nonlocal factor $\Gamma$. In the perturbative string
approach the value of $\Delta$ can be small $\sim g_s^2$, where
$g_s$ is the string constant, if the ``cutoff'' in $\kp$ is on the
string scale and not on the Planck scale, and the higher string
modes entering loops will not compensate this. But if we include
in cascading all other (not only ladder) diagrams, the value of
$\Delta$ can be turn out to zero. This follows from the momentum
sum rules for the parton distribution $n^{\bot}$. The additional
arguments in favor of the condition for a final $\Delta = 0$ are
given in the end of section 4}
at low $\omega$ than the primary WW spectra (\ref{grsp}), and the
value of $\Delta$ gives shift of bare WW intercept~(~=3).

The other important higher order corrections, besides the parton
cascading and various rescatterings (like eikonalization),
correspond to the parton recombination which become essential
when the parton density is high, and this process can lead to the
parton density saturation.

Due to the masslessnes of gravitons there is also the `infrared'
contribution, coming from small $k_{i\bot} \rightarrow 0$. In the
first WW approximation this corresponds to the large impact
parameter elastic scattering. If some cells, in hard ladder,
contain small $k_{i\bot} \rightarrow 0$ than in the impact
parameter space this will correspond to big $\delta b \sim
k_{i\bot}^{-1}$ steps, ``transporting''  part of the hard ladder
far in $b$. In terms of t-channel regge amplitudes this
corresponds to the $\kp^2 \ln \kp^2$ type singularities, coming
from such infrared cells.

Firstly we illustrate this in terms of the parton cascading.
Suppose that the first step (cell) is infrared and consider it.
Then the parton distribution (\ref{grsp}) at large transverse
distances $\xp \gg m_p^{-1}$ can be represented in the form
\bel{dndxdk}
\frac{\d n}{\d\omega\d^2\xp\d^2\xp} ~\sim~
\Big(\frac{E}{m_p}\Big)^2  \frac{1}{\omega^3}~ \tilde{\delta}
\Big(\xp^2 - \frac{1}{\kp^2}  \Big) ~\sim~
\Big(\frac{E}{m_p}\Big)^2 \frac{1}{\omega^3}~ \frac{1}{\xp^4}
\tilde{\delta} \Big(\kp^2 -\frac{1}{\xp^2}  \Big)~, \ee
where $\tilde{\delta}$ is narrow $\delta$-like function, having
width of peak $\sim 1$. Integrating (\ref{dndxdk}) over $\kp$ we
become the distribution of these partons in the transverse (impact
parameter) plane $\xp$
\bel{xsp}
\frac{\d^3n}{\d\omega\d^2\xp} ~\sim~
\Big(\frac{E}{m_p}\Big)^2 \frac{1}{\omega^3}~ \frac{1}{\xp^4}~~,
\ee
which we use for $\xp \gg m_p^{-1}$.  All this corresponds simply
to the approximation $\d n/\d \xp \sim \d n /\d \kp^{-1}$. But if
we insert two or more infrared cells to the cascade chain
(\ref{convol}), we come to a much smaller contribution at large
$b$. Indeed, in this case we have the convolution of two
distributions (\ref{xsp}) containing integrals
$$
    \int\limits_{|\xp -b_i|>m_p^{-1}}  d^2 \xp/(b_1-\xp)^4
         (\xp -b_2)^4 ~\sim~ m_p/(b_1 - b_2)^4
$$
which for $(b_1 - b_2)^2 \gg m_p^{-2}$ gets the main contribution
near the integration ends, where one of these cells enters in the
hard regime. The distribution (\ref{xsp}) at $\xp \gg m_p^{-1}$
refers to soft gravitons,  but inserted in the cascading chain
these partons are the sources for the next hard chain sections.
Then the simple generalization of (\ref{convol}) and (\ref{xsp})
gives the $\xp$ distribution
\bel{xhsp}
 \frac{\d^3n}{\d\omega\d^2\xp} ~\sim~
\Big(\frac{E}{m_p}\Big)^2 \Big(\frac{E}{\omega} \Big)^{\Delta}
\frac{1}{\omega^3}~ \frac{1}{\xp^4}~~,~~~~~ \frac{\d n}{\d\xp^2}
~\sim~ \Big(\frac{E}{m_p}\Big)^{2+\Delta} \frac{m_p^2~}{(m_p
~\xp)^4}~
\ee
for hard partons with $<\kp^2> \sim m_p^2$ and $\xp >,\gg
m_p^{-1}$.

Now consider the interaction of a fast particle, with parton
spectrum (\ref{grsp2}), with a pointlike target. The cross-section
is
\bel{s2g}
\sigma_{in} ~=~ \int d\omega d^2\kp ~n(E,\omega,\kp)~
\hat{\sigma}(\omega,\kp) ~\sim~ s^{2+\Delta}~,
\ee
where $\hat{\sigma}$ is the parton cross-section on a local
target. Such a behaviour corresponds to a regge pole with the
positive signature in a vacuum channel and with the intercept
\bel{intercept}
      \alpha (0) ~=~ 3 ~+~ \Delta~.
\ee
This regge pole can be considered as an analog of the pomeron for
the gravitational interaction
\footnote{It seems that there is no such state between the
perturbative states in known sectors of string theories
containing the gravity. But it can be a composite state seen only
at the strong coupling. It is possible as well that this state is
masked by the singularity corresponding directly to a black disk
(some brane-like object~?).}
(later, for abbreviation, we will call this reggeon 2G). The same
cross-section (\ref{s2g}), written in the impact parameter space,
looks like
\bel{sin-eb}
\sigma_{in}(s,b)   ~\sim~  \Big( \frac{s}{m_p^2}
\Big)^{2+\Delta} ~\frac{1}{(b m_p)^4}
\ee
The singularities of (\ref{sin-eb}) in $t = -q_{\bot}^2$ can be
represented by the following expression
\bel{qsingul}
 e^{-\alpha' q_{\bot}^2 \ln s } ~\Big( a_0 ~+~ a_1 q_{\bot}^2
 \ln\frac{1}{q_{\bot}^2} ~+~ a_2 \big( q_{\bot}^2
 \ln\frac{1}{q_{\bot}^2} \big)^2 ~+...~~ \Big)~,
\ee
giving the main factors for an associated t-channel amplitude,
where terms with coefficients $a_i$ correspond to a contribution
with $i$ infrared cells (loops) in the parton cascade. These terms
lead to the $b \gg m_p^{-1}$ behavior generalizing~(\ref{sin-eb}):
\beal{sin-all}
\sigma_{in}(s,b)   ~\sim~ \Big( \frac{s}{m_p^2}
\Big)^{2+\Delta} \Big(~ \frac{a_0}{\ln s} ~\exp{\Big( \frac{
-b^2}{4\alpha' \ln s}\Big) } ~+~     \nn  \\
~+~~\frac{a_1}{(b m_p)^4}  ~+~
\frac{\tilde{a}_2 \ln b}{(b m_p)^6} ~+... ~~\Big)
\eea
We see that at large $b$ only terms with one infrared cell are
essential. And thus terms in (\ref{qsingul}), singular in
$q_{\bot}$, can be included in vertices connected with the
exchange by the 2G reggeon, with the intercept (\ref{intercept}).
On the contrary, in massive theories like the QCD, only a first
term in (\ref{sin-all}) is present at all $b$ (this is a standard
regge pole contribution). As a result this leads to a fast cutoff
of $\sigma_{in}(s,b)$ in $b$ and, after the unitarization, to the
Froissart-like behavior $\sim \ln^2 s$ of cross-sections.

The cross-sections $\sigma_{in}(s,b)$ grow fast with $E$ and for
some $E$ exceed the maximal value $=1$ allowed by unitarity, when
all incoming particles flux at given $b$ is absorbed. This means
that the parton screening becomes essential. The simplest way to
take screening corrections into account is to use the
eikonalization~:
\bel{eicon} \sigma_{in}(s,b) ~=~ 1 - |\!~S(s,b)\!~|^2 ~=~ 1
~-~\exp \big(- 2\sigma_{in}^0(s,b)~ \big)~, \ee
where $\sigma_{in}^0$ is given by (\ref{sin-all}). The $S$-matrix,
entering (\ref{eicon}), contains, among others, a ``black disk'',
where the parton density exceeds the ``critical'' one $\sim
m_p^{-2}$. This is in the internal part of the AS disk where
\bel{rad}
   \xp^2 ~<~ R_{\bot}^2(E) ~=~ m_p^{-2}
      ~\Big(\frac{E}{m_p}\Big)^{1+\Delta/2}
\ee
Outside this `black' part where $\xp >R_{\bot}(E)$ the density
decreases as~:
\bel{xspx2}
\frac{\d n}{\d\xp^2} ~=~
m_p^2~ \Big(\frac{R_{\bot}(E)}{\xp}\Big)^4
\ee

\nin For $b < R_{\bot}^2(E)$ the screening of partons in a process
of the interaction is essential and the expression (\ref{eicon})
takes it into account. But, at the same time, it corresponds to
that all these partons are nevertheless present in the incoming
state. And therefore, there 2~D density (in parton state) anyway
highly exceeds the Planck density $m_p^{-2}$. But then the mean
transverse momenta of these partons, due to their strong
interactions, also can be much greater than the Planck scale. And
this is inconsistent with our previous assumption that all $\kp$,
much higher than $m_p$, are cut. Therefore we also suppose that
the parton density saturates at the Planck scale and does not
increase with $E$ in an internal parts of the AS disk, where $\xp
< R_{\bot}(E)$. The dynamical mechanism responsible for such a
density stabilization (saturation) can be probably represented as
a parton recombination, like that in the vector (BFKL) case
\footnote{ There are signs that the string system  undergoes some
phase transition at Hagedorn temperatures to some new phase, and
there are arguments \cite{at-wit} that this phase is not a string
like. For the high string densities the situation is probably the
same.~~ And more, there exists a popular opinion formulated in
various forms that virtual (parton) string states with the
transplanckian density of degrees of freedom $\gg m_p^{-D}$ (even
local) are inaccessible, because their contribution is strongly
damped (possibly exponentially in density) in the wave function.~
Despite the fact that such a reduction of high density states can
be encoded in very general constructions, like holographic
principle or M-theory, the concrete mechanism doing this in all
cases can be simply a more fast recombination of additional
degrees of freedom than their creation, when some critical density
is reached.}.

The full $S$-matrix corresponding to (\ref{eicon})
\bel{sma}
   S(y,b) ~=~ e^{ ~i\delta_R(y,b) - \delta_I(y,b) }~~,~~~
   \delta_I(y,b) ~\sim~ \sigma_{in}^0(s,b) ~+~ \cdots
\ee
contains also a real part of the phase $\delta_R(y,b)$, which for
very large $b$ coincides with the ``classical'' AS phase
$$
    \delta_R(y,b)  ~\sim~ \Big( \frac{s}{m_p^2} \Big)~\ln
    \frac{B_0}{b}
$$
and is responsible for a large $b$ elastic scattering. This
S-matrix gives the hard inelastic cross-section
\bel{cin}
\sigma_{in} ~=~ \int d^2b~ (1-e^{-\delta_I(b,y)})
~\simeq~
 \pi R_{\bot}^2(y) ~\sim~ m_p^{-2}\Big(\frac{s}{m_p^2}\Big)~,
\ee
corresponding to the same structure of a black absorbing disk
with the radius $R_{\bot}(y) \sim \sqrt{s}/m_p^2$ as given by the
expression (\ref{rad}).~ Such a disk remains a Froissart black
disk and the method by which it is often introduced from the
eikonalization of supercritical pomeron exchange. The difference
is the behaviour of the black disk radius with the energy~:~
$R\sim m^{-1} \ln (s/m^2)$ for the Froissart case  (for massive
theories),~ and $R\sim m_p^{-2}\sqrt{s}$ for the massless
gravitation
\footnote{There is an interesting question, concerning the
transparency of this black disk, filled with hard gravitons at a
saturated density, and of the dependence of this transparency
from energy. At first it seems that for the saturated parton
disk, with the finite 2-dimensional density of degrees of freedom,
the transparency ($ =~ |S(y,b)|^2$) should remain finite (grey
disk) and not $\rightarrow 0$ when $E \rightarrow \infty$. But
this contradicts probably to the longitudinal boost invariance of
the transparency at a given impact parameter calculated in the
parton model. The same problem emerges in the QCD when we try to
estimate the transparency of the saturated  Froissart disk
\cite{kanch}. But there, with the growth of energy, more higher
transverse momenta scales enter the game. And although on every
such scale saturated density is finite, the total density grows
with $E$ and the disk becomes more and more black. How this
question is solved for gravity is unclear.}.

We considered above only partons with $\omega \sim m_p$~, where
the majority of hard partons are concentrated. We can extend this
to partons with different $\omega$ and find their distributions.
Simple generalization of it gives the parton spectra at arbitrary
$\omega$ and $\xp$
\bel{invsp}
  dn ~=~ \hat{n}(E,\omega,\xp)~(m_p^2~ \! d^2\xp)~
  \frac{d\omega}{\omega} ~~,~~~~~\hat{n} = \Big(\frac{E}{\omega}
  \Big)^{2+\Delta} \frac{1}{(\xp m_p)^4}~~,
\ee
where the density $\hat{n}$ is dimensionless and boost invariant.
This distribution takes an even more simple form if we go to
rapidities $Y=\ln E/m_p,~ y=\ln \omega/m_p$ and to $\zeta = \ln
(\xp m_p)^2$~:
\bel{invspy}
dn ~=~ e^{(2+\Delta)(Y-y)-\zeta} ~dy~ d\zeta
\ee
From the condition $\hat{n} ~\sim~ 1$ we find the black disk
radius for various $\omega$~:
\bel{radom}
 R_{\bot}^2(E,\omega) ~=~ m_p^{-2}
   ~\Big(\frac{E}{\omega}\Big)^{1+\Delta/2}
\ee
Thus the saturated part of a hard parton configuration in $(\zeta,
y )$ space consists from a ``cone'', with the boundary $\zeta <
(2+\Delta)(Y-y)$ ~(this corresponds to $\xp <
R_{\bot}^2(E,\omega)$), filled with the density $\sim 1$ ~($\sim
m_p^{-3}$ in units $\xp$). This cone is surrounded by a more
``diffuse'' soft ($\omega \ll m_p $) cloud, where the parton
distribution is given directly by (\ref{invspy}).

\nin For $\omega \ll m_p$ the parton density $\hat{n}$ in this
soft tail can be very large $\hat{n} \gg 1$. The mean transverse
localization of these partons is $\delta \xp \sim 1/\omega$.
Comparing it with the size of the region $R_{\bot}(E,\omega)$
where $\hat{n} > 1$ we can find the soft critical radius and the
parton frequency
$$
\tilde{R}(E) \sim   m_p^{-1} \Big( \frac{E}{m_p}
  \Big)^{1 + 2\Delta/(2-\Delta)}  ~,~~~~~~~~
  \tilde{\omega} \sim 1/\tilde{R}(E)~,
$$
which define the border of the fully coherent parton region.  For
$\xp > \tilde{R}(E)$ the parton interaction with the target is
week and is represented by the ``classical'' elastic scattering
and the soft classical gravitational wave emission.

\section{\bf Collision of two AS disks}

Firstly, let us consider an interaction of two transplanckian
particles, ``represented'' by such parton disks, in a formal way
- by the regge like machinery. Usually in this case one starts
from a Green function corresponding to a basic Regge object (like
bare pomeron) and then, using it, one can build various higher
order diagrams for amplitudes and inclusive cross-sections. In
our case this implies to start from 2G pole with
$\alpha(0)=3+\Delta$ corresponding to (\ref{s2g}). But we make
one step forward and start directly from the ``economized''  2G
object. For pomerons this step corresponds to a transition to
``Froissarons'', which are the economized sums of pomerons, and
then to construction of higher diagrams using such objects
\cite{cardy}.~ Applying this approach to our case we form the
basic reggeon Green function - proportional to the amplitude
corresponding to the S-matrix (\ref{sma})~:
\bel{green}
   D(y,b) ~=~ i(1 - S(y,b)) ~\sim~ i~\theta \big(R_{\bot}^2(y)
   - b^2\big) ~+~ D^{soft}(y,b)~,
\ee
where $D^{soft}(y,b)$ is the soft part of $D$ containing
gravitons with $\omega,~ \kp \ll m_p$. Using this $D(y,b)$ one can
construct higher reggeon diagrams and various inclusive
cross-sections. The hard inelastic cross section is given by
\bel{siinel}
 \sigma_{in}(Y) ~=~ \int d^2B~ \hbox{Im} D(Y,B) ~\simeq~
  \pi R_{\bot}^2(Y)  ~~+~  \sigma_{in}^{soft}(Y)
\ee
as considered in the previous section. The corresponding inclusive
cross-section for hard gravitons with the definite impact
parameter $b$ and the rapidity $y$ and the initial $Y$ and $B$ is
given by the reggeon diagram with two cut D~s ~-~ it takes in the
impact parameter representation a simple form
\beal{hinc}
 \rho(y,b,Y,B) ~\simeq~ V~\hbox{Im} D(y,b)\cdot
  \hbox{Im} D(Y-y,B-b)~~=~~~~~~~~~~~~~~~~~~  \\
=~ V~\theta \big(R_{\bot}^2(y) - b^2\big) \cdot
 \theta \big(R_{\bot}^2(Y-y) - (B-b)^2\big)
 ~+~ \rho^{soft}(y,b,Y,B)~~, \nn
\eea
where $V$ - is the inclusive graviton emission vertex, and the
$\rho^{soft}$ - the soft contribution. Other higher inclusive
cross sections can be constructed in the similar way. The region
where such $\rho(y,b,Y,B)$ is big is shown in Fig.4.
\begin{figure}[h]    \centering
\begin{picture}(80,125)(130,-10)    
\includegraphics*[scale=.5]{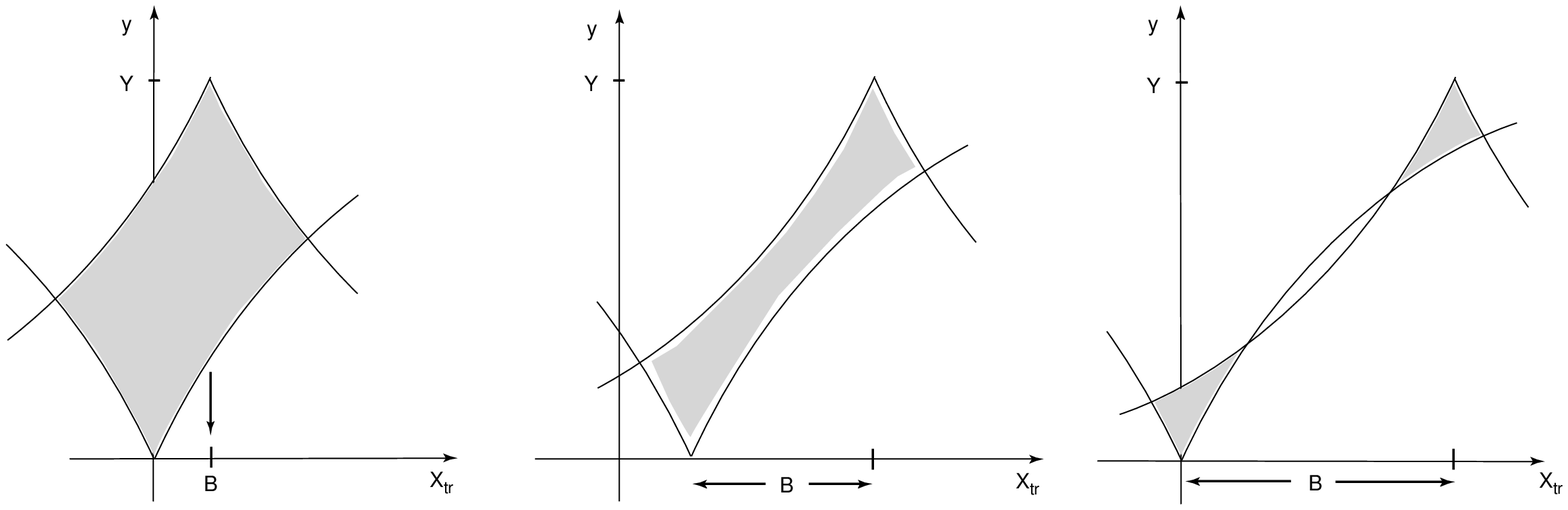}
\end{picture}
\parbox{10cm}{{\bf Fig.4~:}~~ The shape of region of intersection
of black disks in $(\xp ~*~ y)$ space at various B. The width of
shaded region at given rapidity is connected with value of
inclusive cross-section as represented in Fig.6}
\end{figure}
The~higher loop corrections in $D$ do not change the structure of
the inclusive cross-section in the ``black'' part of $D(y,b)$ i.e.
at $b < R_{\bot}(y)$, but can be essential at the border of the
AS-disk, where their influence on a diffractive and a soft
radiation processes (like in the case of Froissaron \cite{cardy}
) is not a small.

Now let us try to interpret the expressions (\ref{hinc}) in the
parton language. In terms of multiparticle Fock space components
this is rather a complicated task  and is not yet considered in
full details even for a simple scalar multiperipheral interaction.
But we know many qualitative aspects of such processes from the
space-time interpretation of various ladder and eikonal diagrams
corresponding to a reggeon exchange. We consider separately soft
and hard gravitons. Because most of partons in the AS disk have
energies $\omega \sim m_p$ we will consider firstly only them.

\vspace{4mm} {\nin  \bf \textit{Hard graviton production.}}

\nin  At given rapidity $y$ and $\vec{x}_{\bot}$ hard gravitons
can be produced only from a collision at the same rapidity and
$\vec{x}_{\bot}$ of black components of two disks. This
corresponds directly to the ``two $\theta$'' intersection term in
(\ref{hinc}). Then the full inclusive spectrum at the same $y$ is
proportional to the area of the intersection of two black parts of
AS disks at the given rapidity. So, for definite $B$ and $y$, we
can have three different configuration of disks intersections~ as
shown in Fig.5.
\begin{figure}[h]    \centering
\begin{picture}(70,100)(80,-10)
\includegraphics*[scale=.4]{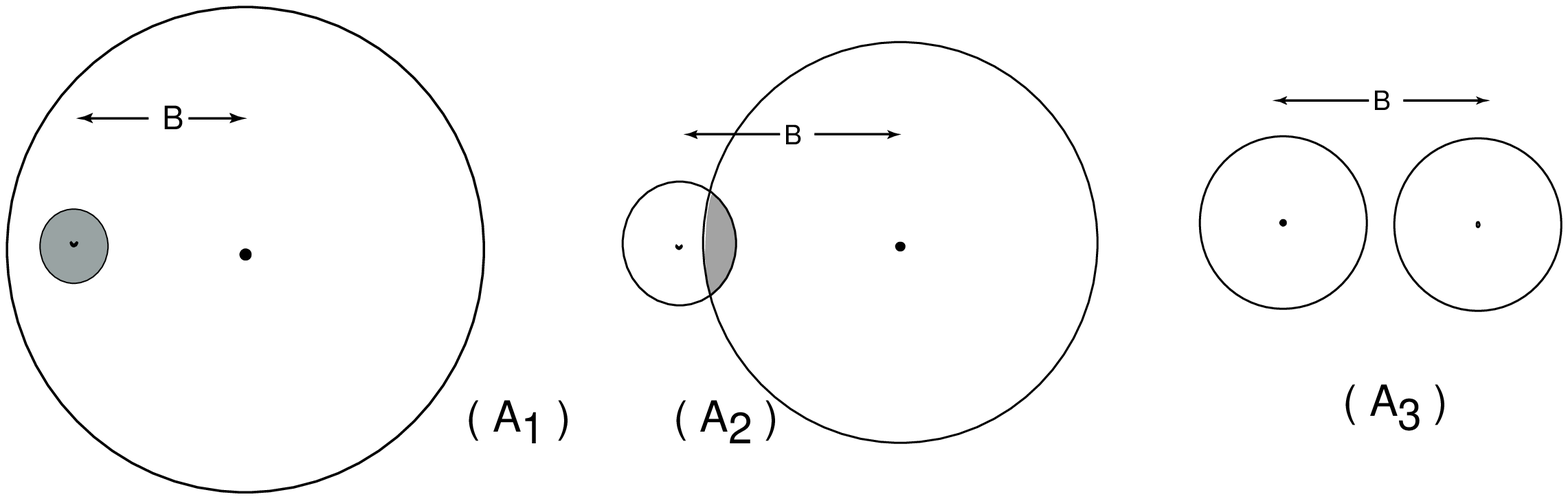}
\end{picture}
\parbox{10cm}{{\bf Fig.5~:}~~
Configuration of intersecting black AS disks is sown at three
rapidities: $A_1$ - closer to the fragmentation region, $A_3$ - in
the center of mass system. This can be compared  with spectra at
Fig.6 at  $B > B_{cr}$ }
\end{figure}
In the cases ($A_1$) and ($A_2$) the two AS-disks intersect - and
the area of the intersection, in Planck units,
$A(B,R_{\bot}(E_1),R_{\bot}(E_2))$ defines the value of hard
inclusive cross-section. In the configuration (b) there is no
intersection of the black parts of the disk - so the contribution
to the hard inclusive cross-section at these $B$ and $y$ is zero.
We neglect effects of smeared hard disk borders and then the hard
inclusive density at given $B$ and $y$ is approximately given by
\bel{roin}
 \rho(B,y,Y) ~\simeq~ m_p^2~A(B,R_1,R_2) ,
\ee
where $R_1=R_{\bot}(E_1),~R_1=R_{\bot}(E_1),~~ E_1 = m_p
e^y$,~$E_2 = m_p e^{Y-y}$, and the area of the black disk
intersection can be represented as
\bel{area}
 A ~\simeq~ \int d^2 \xp~\theta(R_1^2-\xp^2)
 ~\theta(R_2^2-|\xp-B|^2)~~\Rightarrow~~~~
\ee
$$
A ~=~ \zeta_1 R_1^2 ~+~ \zeta_2 R_2^2
  ~~~~~~~~~\hbox{for}~~~~~~  |R_1-R_2| < B < R_1+R_2~,~~
$$
$$
 A ~=~ \min (\pi R_1^2,~ \pi R_2^2)~~~~~~\hbox{for}~~~~~~
 B < |R_1-R_2| ~,~~~~~~~~~~~~~~~~
$$
where
$$
\zeta_1 ~= \theta_1- \frac{1}{2}\sin 2 \theta_1~,~~~~
\zeta_2 ~= \theta_2- \frac{1}{2}\sin 2 \theta_2~,~~~~~
$$
$$
  \cos \theta_1 ~=~ \frac{B^2 + R_1^2 -R_2^2}{2 B  R_1}~,~~~~~
  \cos \theta_2 ~=~ \frac{B^2 + R_2^2 -R_2^2}{2 B  R_2}
$$
The structure of hard graviton spectra in rapidity, corresponding
to (\ref{roin},\ref{area}), is illustrated at Fig.6~
\footnote{Later on we put $\Delta =0$. This simplifies expressions
and at the same time corresponds probably to the correct value.}.
It changes from highly peaked in c.m.s. for $B\sim 0$ to the
purely diffractive like spectra with two filled bumps near the
fragmentation region of colliding particles for large $B \sim
m^{-2}_p \sqrt{s}$.
%
\begin{figure}[h]    \centering
\begin{picture}(70,170)(80,-10)
\includegraphics*[scale=.45]{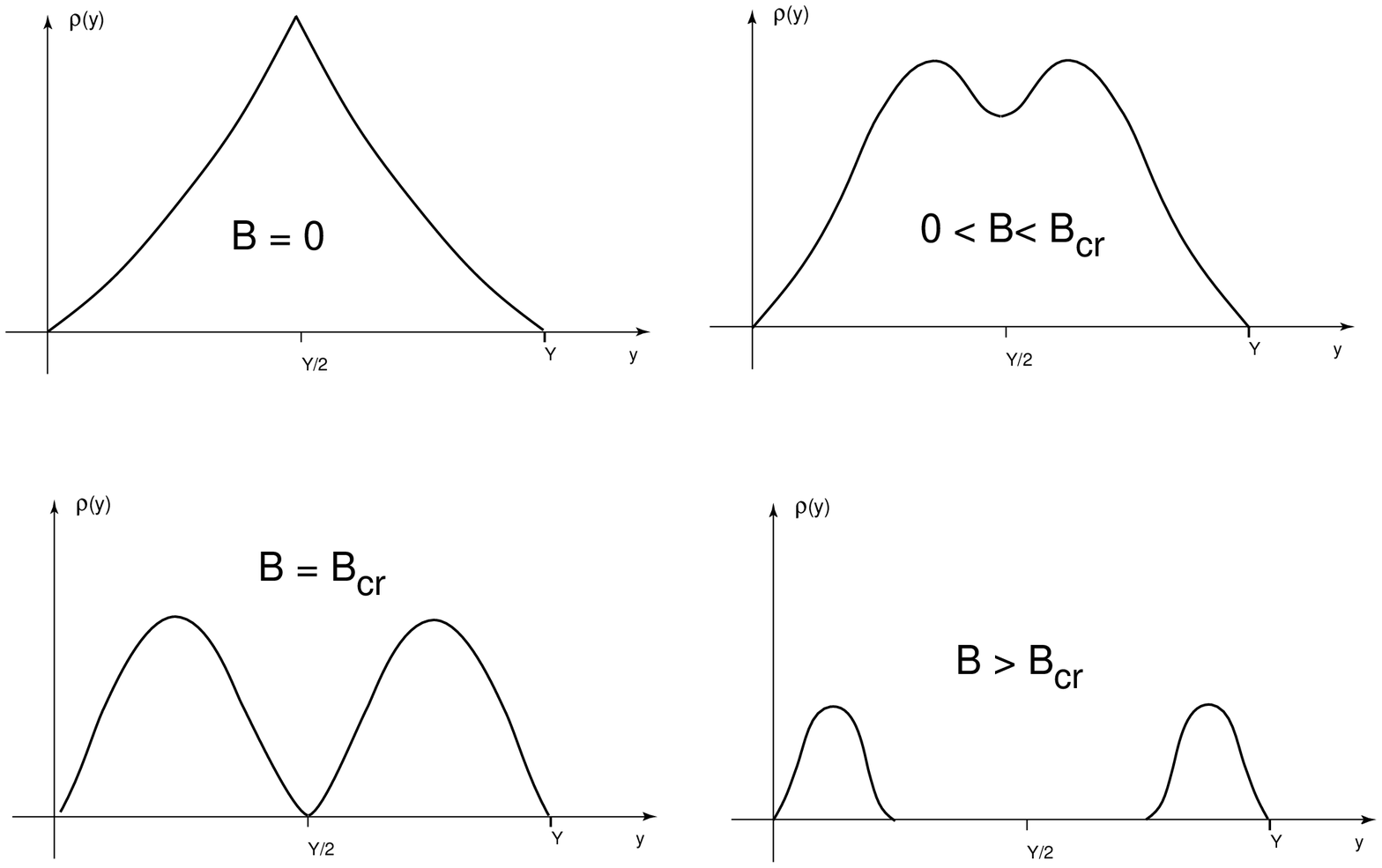}
\end{picture}
\parbox{10cm}{{\bf Fig.6~:}~~ Hard gravitons inclusive spectra
in rapidity (a,b,c,d) for various values of the impact parameter
$B$, where $ B_{cr} = 2 R_{\bot}(\sqrt{s} /2 ) $. }
\end{figure}
The gap in inclusive spectra around the central position $y=Y/2$
is formed when $B > B_{cr}$, where the $B_{cr}$ correspond to a
configuration when two hard disk in c.m.s. only touch each other,
i.e. $2R_{\bot}(Y/2) = B_{cr}$. The multiplicity of produced
gravitons and the maximal value of the spectra at Fig.6 also
depends strongly from $B$, and drastically changes their behaviour
at ~$B \sim B_{cr} \sim m_p^{-1}(s/m_p^2)^{1/4}$.~ From
(\ref{area}) one can simply estimate the behaviour of the mean
multiplicity of hard gravitons as a function of $s$ and $B$:
\bel{mul}
N(B,s) ~=~
\theta\big(B_{cr}-B\big)~\frac{\sqrt{s}}{m_p}
~f\Big(\frac{B}{B_{cr}(s)}\Big)
~~+~ \theta\big(B-B_{cr}\big)~\frac{s}{m_p^4B^2}~,
\ee
where the function $f(B/B_{cr}) = m_p s^{-1/2}\int dy A\big(
B,R_{\bot}(y),R_{\bot}(Y-y) \big)$ depends slowly on the argument.
The multiplicity distribution due to (\ref{area}) is also mainly
defined by geometrical parameters of collision, and the
probability to produce $n$ hard gravitons
\bel{mudlis}
 w_n ~\simeq~ \frac{1}{n^3}~+~
 \Big(\frac{m_p^2}{s}\Big)^{-3/2}
 ~f_1\big(n \frac{m_p}{\sqrt{s}}\big) \theta(n-n_{cr})
\ee
is fast decreasing almost for all $n$. The small additional term
with the slowly varying function $f_1 \sim 1$ in r.h.s. of
(\ref{mudlis}) is essential only for $n>n_{cr} \sim \sqrt{s}/m_p$.
It comes from collisions with $B<B_{cr}(s)$. Although the total
inelastic cross-section (\ref{siinel}) for the production of hard
gravitons is big $\sim s m_p^{-4}$~, the corresponding final state
contains a small number of particles (Fig.6~d).~ The processes
~(Fig.6 a,b,c)~, where the hard multiplicity in final state is
high $\sim \sqrt{s}$, come only from collisions with $B < B_{cr}$
~-~ their cross-section is $\sim m_p^{-3/2}\sqrt{s}$.

The mean hard graviton multiplicity corresponding to
(\ref{mudlis}) is low $~<n> \sim 1$ at all $s$, and the
dispersion of the multiplicity grows: $ <n-<n>>^2 \sim \ln s $,
but not so fast
\footnote{This radically differs from possible expectations, when
for all $B < m_p^{-2}\sqrt{s}$~ the big black hole (possibly
rotating) is created, leading at the end to a almost full
dissipation of the initial energy into soft in c.m.s. particles.
}.

Note that as a result of such behaviour of $w_n$ the full
inclusive spectra of hard gravitons integrated over $B$ takes a
very simple scaling form
\beal{fin}
\tilde{\rho}(y,Y) = \sigma_{in}^{-1}(Y)
 \int d^2B~\rho(B,y,Y) ~\simeq~~~~~~~~~~~ \\
 \simeq~ \frac{\pi m_p^2~R_1^2~R_2^2}{R^2(s)} ~=~
 \pi\frac{E_1\cdot E_2}{m_p E} ~\sim~ const(y,Y) \sim 1  \nn
\eea
like in the case of constant asymptotic cross-sections. Although
here the fluctuations in shape of the individual events can be
very big and are mainly governed by the distribution of impact
parameters, like in the case of a nucleus-nucleus collision.

\vspace{4mm} {\nin  \bf \textit{The role of soft gravitons and the
longitudinal boost behaviour.}}

\nin We have not considered explicitly the production of low
energy gravitons with $\omega \ll m_p$ and of the soft gravitons
with $\kp \ll m_p$. There are two types of such processes. In one,
which takes place on large impact parameters, only the soft
gravitons are produced. Such processes must be more or less
satisfactorily described by the perturbative methods - and the
complete space-time picture will be not far from that
corresponding to a classically described gravitational wave
radiation during a fast particles collision, for example like that
given in~\cite{eath2}.

The other processes, taking place at smaller $B$, also contains
soft gravitons accompanying hard gravitons created as a result of
a collision of black AS disks. But here the soft and low energy
partons also enter directly in the production of hard gravitons.
The consideration of their interaction is needed to consistently
explain (in parton terms) the generation of hard particles for
relatively  big impact parameters $B_{cr} < B <
m_p^{-2}\sqrt{s}$~,~ and the inclusive spectra of type Fig.6d,
when we view on a collision from the c.m.s., where the hard disks
do not collide directly.

The related more general question concerns the invariance of
cross-sections and reactions outcomes under longitudinal boost
transformations. The parton state of fast particles changes under
such boosts in a complicated and nontrivial way: the mean number
of partons and their space configuration changes etc. Therefore
the requirement of the invariance of various cross-sections
calculated in the parton approach under boost, corresponds to
very strong condition, essentially restricting the structure of
the parton state itself~
\footnote{In the massive theories like QCD, the rate of the growth
of a parton black disk radius $\Rp (E)$ is almost completely fixed
by the invariance of the inelastic cross-section $\sigma_{in} =
\pi (\Rp (E_1 \xi) + \Rp (E_2 /\xi) )^2$ under changes of the
boost parameter $\xi$. Such a condition leads only to two
solutions~: $\Rp(E) = const(E)$ and $\Rp(E) = c \ln E$.~ This
corresponds either to $\sigma_{in}(E) =const$ or $\sigma_{in}(E)
\sim \ln^2 E$~ - that is the Froissart-like behaviour. Moreover,
by same method one can show that in the Froissart case only the
black (not a gray) disk case is allowed. The other restriction
can also be found by the same way. The reason why the bust
invariance of cross-sections is so restrictive for parton states
is probably connected with that it plays the role of t-unitarity
which is very essential for high energy amplitudes, but can not
be explicitly imposed on parton wave functions \cite{kanch}.}.

Let's consider the interaction of the AS disk with $E_1 \gg m_p$
with an almost massless particle which can have a ``low'' energy
$E_2 \ll m_p$ - so that it can be considered as ``test'' objects
and in the same way choose the full $s \simeq 2 E_1 E_2 \gg
m_p^2$.

Then examine the behavior of this system under the longitudinal
boost - when the transformed energies become
\bel{bt}
     E_1 ~\rightarrow~ E_1~ \xi~,~~~~ E_2 ~\rightarrow~ E_2 / \xi~,
\ee
but the impact parameters $B$ and $s$ are not changed.

Firstly choose $B = B_1 < R_{\bot}(s) \sim
m_p^{-1}(s/m_p^2)^{1/4}$ so that the hard gravitons are produced
in the final state. If $\xi$ are such that $E_2 \sim m_p$ then we
have the collision of the $AS_1$ black disk with a hard particle,
which is inelastically scattered by a disk with the probability
$= 1$. After that the production of secondary hard gravitons
starts almost immediately and the real particles appear at times
$\sim m_p^{-1}$. Call this frame - frame I and fix $\xi \sim 1$
in this frame.

If we choose next $\xi$ such that $E_2 \ll m_p$ then we must
suppose that the interaction of the ``$E_2$'' particle with the
hard saturated $AS_1$ disk at the same $B$ takes place with the
same probability $=1$, although the cross-section of the
interaction with individual partons decreased $\sim 1/\xi^2$. Now
the $AS_1$ parton system is only ``softly'' excited during the
interaction. After that the instabilities in the $AS_1$ disk grow
gradually and first hard gravitons will be created only after the
AS disk moves in $z$ on $\sim m_p^{-1}~\xi$ from the collision
place. Let's call this frame - frame II.

But for such $\xi$ ~(in frame II) the radius of the black part of
the $AS_1$ disk, where partons are in the saturated phase, is also
increased
$$
   R_{\bot}(E_1) ~\rightarrow~ R_{\bot}(E_1 \xi) ~=~
   \xi^{1/2} R_{\bot}(E_1)~.
$$
Therefore the ``$E_2$'' particle should be absorbed by a disk with
the probability $= 1$ for larger $B$ up to $\xi^{1/2}
R_{\bot}(E_1)$. This  probability of the particle ``$E_2$''
interact at $B=B_2$ where $R_{\bot}(E_1) < B_2 < R_{\bot}(E_1
\xi)$ must not depend on $\xi$. In the frame I at $B=B_2$ the
``$E_2$'' particle does not collide directly with a black disk.
Now directly with the black disk can interact only soft partons
accompanying the particle ``$E_2$''. And this mechanism can
regulate the boost invariance of the probability interaction $=1$.
This role of soft partons-gravitons is even more evident if, at
the same $B=B_2$, we select $\xi$ so to move to c.m. system, where
the black disks are most far one from another in the transverse
direction.

One can estimate simply maximal impact parameters for which the
capture of soft partons from one colliding particle by the black
disk of another particle take place with the probability = 1. For
the capture of soft partons we need to choose $B$ and $\omega$ so
that as minimum as one soft parton from ``$AS_2$'' falls into the
black disk of $AS_1$ or vice versa - this gives~:
$$
\hat{n}(E_2,\omega,B) \cdot \Rp^2 (E_1,\omega_1 \sim m_p) ~\sim~
m_p^{-2}
$$
From the other hand, the size of the transverse localization of
the soft parton, which is $\sim~1/\omega$, must be much less than
the $AS_1$ black disk radius $\Rp (E_1,\omega_1 \sim m_p)$.
Otherwise the soft parton will mainly elastically scatter on the
black disk and not be captured by him. This corresponds to the
condition for minimal $\omega$~:
$$
    \omega \cdot \Rp (E_1, m_p ) ~\sim~ 1
$$
Then, using expressions (\ref{invsp}),(\ref{radom}) for $\hat{n}$
and $\Rp$~, it is simple to find $B = R_{\bot max}$ and $\omega$
fulfilling these two conditions in the case of the arbitrary
intercept~$\Delta$~:
$$
  \omega_{min} ~\sim~ m_p \Big( \frac{m_p}{E_1} \Big)^{(2+\Delta)/4}
      ~~,~~~~~~ (B m_p)^2 ~\sim~ \Big( \frac{E_1}{m_p}
      \Big)^{1+\frac{3}{4}\Delta+\frac{1}{8}\Delta^2}
      ~\Big( \frac{E_2}{m_p} \Big)^{1+\Delta/2}
$$
The inelastic processes up to such big impact parameters $B \sim
R_{\bot max}$ are essentially influenced by ``strong'' gravity
mechanisms. The corresponding hard inelastic cross-section
\bel{si-xi}
   \sigma_{in}(E_1, E_2) ~\sim~
    (R_{\bot max})^2     ~\sim~ m_p^{-2}~\Big( \frac{E_1}{m_p}
    \Big)^{1+\frac{3}{4}\Delta+\frac{1}{8}\Delta^2}
    ~\Big( \frac{E_2}{m_p} \Big)^{1+\Delta/2}
\ee
can be boost invariant under (\ref{bt}) only if the powers of
$E_1$ and $E_2$ in (\ref{si-xi}) are equal - because only in this
case $2E_1 E_2 \sim s$ does not depend on $\xi$. This gives
condition on $\Delta$ with two solutions
$$
     \Delta ~=~ 0 ~,~~~~~~ \Delta ~=~ -2
$$
The first solution corresponds to the hard $ \sigma_{in} \sim s
m_p^{-4}$ cross-section which we have discussed above and gives
the additional arguments for the $\Delta =0$ case. The second
solution corresponds (a little peculiarly) to asymptotically
constant cross-sections, typical for hadron interactions and
associated with a pomeron.

\section{\bf  ~~Instability of the final state.\\
The black holes creation after collision}

Before the collision, in parton black disks of incoming particles,
we already have too much `low' energy transverse partons-gravitons
so that in fact the full virtual energy is concentrated in them.
But this state is evidently stable under the implosive collapse
due to strong phase correlations between these partons. During
the collision some partons are exited and ``discorrelated'', and
later become free with the inclusive spectra of type
(\ref{hinc}). Only after that the possible long-range
gravitational instabilities in the evolved particles system can
lead to the formation of various density clusters over the
``initially smooth'' $\rho(y)$ background.

The main attractive instability can take place for groups of
created particles having close longitudinal velocities.
We can divide conditionally the $\rho(y)$ spectrum in layers in
the rapidity of the width $\delta \sim 1$ in such a way that the
relative energies of particles(gravitons) from the same layer are
not high, and the hard partons from the neighbor layers look
ultrarelativistic. In a first moment the created particles
density in every such layer is high - close to the critical.
Therefore these particles (gravitons) strongly interact one with
another during some period of time before there density becomes
so small that they can move almost free. During this period of
time the relative energies of particles in layer are partially
thermalized and they behave in fact as a massive particles. It is
essential that because of such multipole scattering on ``media'',
for some period of time particles can not escape from layer as
free massless particles. In fact such condition fixes the value
of $\delta$ and the portion of particles that are ``confined'' in
layer.

If we go to a longitudinal system with a definite rapidity $y_1$,
where some layer is ``standing'' (marked in Fig.7) then we see
that particles from other layers move fast away. In fact in the
corresponding time scale these particles can be still in the
virtual state and become free only after time $t \sim
\exp{|y-y_1|}$.
\begin{figure}[h]    \centering
\begin{picture}(70,110)(50,0)
\includegraphics*[scale=.3]{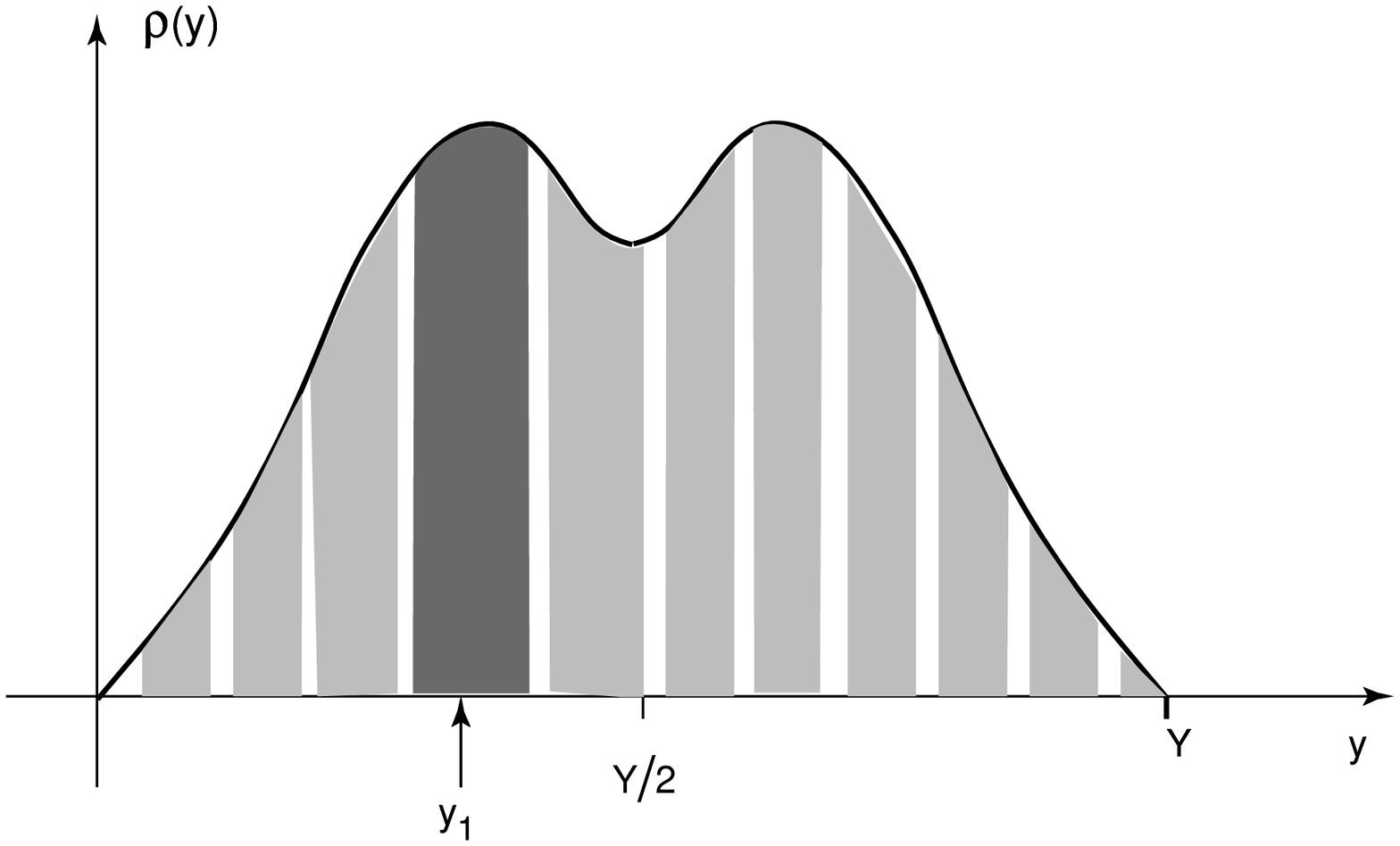}
\end{picture}
\parbox{10cm}{{\bf Fig.7~:}~~
Inclusive spectra for hard gravitons divided on layers in
rapidity. These layers can transform during the following
evolution into black holes with the same rapidities and
masses~$\sim~\rho(y)$.}
\end{figure}
So we can consider this layer separately from the others. Its
mass is
\bel{mass}
 M(y_1) ~\sim~ m_p^3 A(y_1,Y,B) ~\sim~ \rho(y_1,Y,B) ~,
\ee
where A is the transverse area of the ``$y_1$'' layer. This mass
is concentrated in the region of the space with the longitudinal
size $\sim m_p^{-1} \exp{(\delta)}$~ and the transverse size $\sim
\sqrt{A}$. These magnitudes are much less than the size of the
horizon of the BH with the same mass $M(y_1)$, and at the same
time particles constituting this mass are not ultrarelativistic.
Therefore the most part of particles from such a layer at time
$\sim \sqrt{A}$ transforms in a black hole with the mass $M(y_1)$
and the transverse position, defined by the center of the region
$A$. During this time the horizon ``surrounding'' these particles
transforms from a pancake-like configuration to sphere.

The neighbours to ``$y_1$'' layers ``$y_1-\delta$'' and
``$y_1+\delta$'' also transform to black holes with masses $m_p
\rho(y_1-\delta)$ and $m_p \rho(y_1+\delta)$~, but this will take
place later and they will have the relative to ``$y_1$''
velocities $\sim 1$.~ So gradually the chain of black holes will
be produced whose distribution in longitudinal momenta is
typically multiperipheral and close to the uniform
\footnote{In fact uniform distribution in rapidity can be
slightly distorted, because the effective  widths  of layers  in
rapidity can be correlated with there masses.}
- with $\rho = const(y) \sim \delta^{-1}$. But their masses are
different at fixed $B$ and are given by values of
$\rho(y_1,Y,B)$. On average - when we integrate over $B$  then the
mean masses of produced BH are also constant over $y$ and small
$\sim m_p$, because the main contribution comes from the large
$B$.

Because every such BH is produced from a layer of partons having
approximately the same longitudinal rapidity,  these BH will be
non-rotating. This is accurate within the small ``thermal''
fluctuation, and border effects. Because the borders of layers are
slightly asymmetric for  $B \neq 0$~~(see Fig.4) some angular
momentum can be concentrated there. The main part of the angular
momentum from the initial state (especially for large $B$) is
transferred to the relative motion of the created BH and to the
surrounding particles not captured by BH.

The rapidity distribution of produced BH in the individual events
is controlled mainly by the corresponding values of $B$ and
repeats the curves in Fig.6.~~ For $B<B_c = m_p^{-1}(E/m_p)^{1/4}$
we have a continuous (multiperipheral) in $y$ chain of BH, whose
masses are given by curves Fig.6(a-b). For $B>B_c$ two chains of
black holes moving in forward and backward directions are created
- their masses repeat again the behaviour of $\rho(y,Y,B)$ in
Fig.6(c,d). These chains are separated by the rapidity gap of
width $\Delta y \simeq Y - 2\ln(B m_p)$ and for big $\Delta y $
when $B \sim \sqrt{s}/m_p^2$ correspond to a typical diffractive
generation of black holes~
\footnote{
Above we discussed how complicated can be the ``trajectories''
leading from the colliding particles state to a final state
containing the BH. Such trajectories are evidently absent in the
semiclassical probability estimate
$\exp{\big(-S_{gr}(\tilde{g})\big)}$, when we put in Euclidean
classical action  $S_{gr}$ some continued solutions $\tilde{g}$
containing the BH. There are two ways to cure this. In one, we
must put all higher radiative correction in the Euclidean action
(the necessary number grows with the energy). Then the
corresponding equation of motion can have necessary trajectories
$\tilde{g}$ with a low value of effective action and a high
statistical weight. The other way is to use the classical action
and consider the transition to the BH not from the few particle
initial state, but from the specially prepared multiparticle
state (parton component), containing approximately the same
number of degrees of freedom as the entropy of the BH. But in both
cases the ``advantages'' of the Euclidean approach disappear and
the consideration in the Minkowski frame is much simpler.~ This is
in fact the general reason because the Euclidean methods do not
work properly at high energies.}.

Probably the part of particles falling between layers can escape
and not captured by one of BH. Their percentage is possibly not
big, but it is complicated to estimate them by the qualitative
methods used here.

It remains a question~:~ can some of these black holes (or all of
them) merge during the future evolution, if the rate of
longitudinal grow of there horizons is faster than these BS
separate one from another. In fact this is a purely classical GR
problem, and it needs additional investigation. Here we present
only following qualitative argument.

The longitudinal structure of particles system created in a
collision of black AS disks is similar to the one-dimensional
homogeneous expanding cosmological solution. In both cases
particles are ``injected'' initially in such a way that in every
longitudinally boosted frame we locally see particles flying away
with the same 2D density and the momentum distribution.~ Usually
for such expanding system the small long-range density
perturbations are damped with time or not growth. Only a
short-range (scale $\delta$) large density fluctuations, created
on the initial ``viscous'' stage, can growth and clusterize. But
on a scale much larger than this no new structures emerge.

\section{\bf  Conclusion and the final remarks }

Thus we come to a picture of gravitational interactions at $E \gg
m_p$ which is in some respect reminiscent to the picture of
asymptotic hadronic (QCD) interactions in the Froissart limit: the
transplanckian particles collision looks like an interaction of
black disks with the production of particles in the geometrical
intersection region of disks. The essential difference is that
for the gravitational interaction the radiuses of these black
disks grow much faster $\sim E^{1/2}$ with energy, as compared to
the QCD case, where these radiuses grow relatively slow $\sim \ln
E$. The other distinction is the character of the interaction in
the final state. In the QCD the final state hadronization is
short-range and so this does not essentially change the structure
of the final state. But for the gravitational case we have a
long-range attractive interaction which can create a
``multiperipheral'' chain of black holes. Finally all this
difference is the consequence of the masslesness of gravitons. If
QCD were massless at large distances then probably also there can
some similar phenomena originate - like powerlike growth with
energy of inelastic cross-sections, etc.

\vspace{4mm} \nin Additional remarks in conclusion.
\begin{itemize}
\item
At ``not too high'' energies ~$s \sim m_p^2$~, when both black
disks from hard partons are only in the embryonic state,~ it is
complicated to make any quantitative conclusion (in particular -
predict the threshold for the first BH creation and their mass) by
the methods we used. Here all essential parameters are of order
of unity and the possible dynamics can be rather involved. But
there are also no reasons to expect that the purely classical
estimates give reliable values.

\item
This model can probably extended to the case with additional
hidden dimensions $D > 4$.~ Here one can distinguish two cases

a) ``Small'' additional  compact dimensions, with the radius close
to the Planck scale $\sim m_p$. In this case all higher
Kalusa-Klein modes act simply as massive particles strongly
coupled at the scale $\sim m_p$ to gravitons. Probably they do not
change any qualitative aspect of picture, but contribute ``only''
to various renormalization of numerical parameters.

b) If the size of additional dimensions  (or of some of them) in
which gravitons can penetrate is very large compared to the 4~D
Planck scale, as in the case of the TeV gravity \cite{add}, then
some aspects of the picture can change more essentially. In this
case too much depends on the concrete realization of the gravity
in the brane world. For example, in the case of only large and non
warped additional dimensions we simply have the $D>4 $dim.
gravity, but some particles (and the colliding ones also) are
glued to 3 space dimensions. Then we have the transverse graviton
spectrum ~$dn \sim (E^2 d\omega /\omega^3) d^{D-2}(\kp/m_p)$.~
This leads to a black D-2 dimensional disk filled with partons at
the Planck density with the radius
$$
 R(E) ~\sim~ m_p^{-1} \Big( E/m_p \Big)^{1/(D-2)}
$$
and to cross-sections $\sigma_{in} \sim R^2(E)$. By the same way
one can find the created hard graviton density $\rho(y,B,Y \sim
m_p^{D-2}A((B,y,Y))$, where $A$ is now the $D-2$ dim. volume of
the colliding disk intersection region. Qualitatively the B
dependence on $\rho$ is the same as  in Fig.6. In the final state
one can again expect the gravitational long-range instability and
the produced particles clusterization into the chain of black
holes. But now (and here much depends on the details of brane
world models) created gravitons and black holes can move in large
additional dimensions and dissipate there. In corresponding
``accelerator events'' this can look out as if a part of the
colliding energy escapes.~ Evidently there can be many variations
of this simple scenario.

\item
In this model the elastic scattering amplitude contains
components coming from different sources. One is Coulomb like - it
comes from the ``classical'' soft graviton exchange at large
impact parameters. Of same type are the soft long-range
multiperipheral corrections. The other one corresponds to the
imaginary diffraction contribution to the elastic amplitude
generated by the big inelastic cross-section $\sim
m_p^{-2}(s/m_p^2)$.

The finite angle high energy elastic scattering usually comes
from small impact parameters and is triggered by the fluctuations
in which in an initial state one has only few (minimally possible)
number of partons, concentrated in a small size volume. `For
transplanckian collisions this corresponds at least to components
of partonic wave functions of both particles without hard black
disks. This probability is $~\sim \exp (-s/m_p^2~ C)$, and this
leads to the too fast decreasing term in the finite angle
scattering cross-section~:
$$
   d \sigma/d \theta ~\sim~ \frac{s}{m_p^4}
   ~\exp \Big({-\frac{s}{m_p^2} C(\theta)}\Big) ~+~
    \Big( d \sigma/d \theta \Big)_{soft}~,
$$
where $\Big( d \sigma/d \theta \Big)_{soft} \sim \exp (-c_1
\sqrt{s}~)$ is the large impact parameter contribution, generated
by the multiple scattering. This behaviour can be compared with
the Cerelus-Martin bound and with the finite angle cross-section
for a string scattering \cite{gross-mende}.

\end{itemize}
\vspace{5mm} \nin {\bf ACKNOWLEDGMENTS}

I would like to thank
A.B.~Kaidalov and K.A.~Ter-Martirosyan for useful conversations
and comments
and I.N.~Kancheli for help and advices.

A financial support of  RFBR through the grants 01-02-17383 and
00-15-96786 is gratefully   acknowledged. This work was also
supported  by  the  INTAS grant  00-00366 and NATO grant PSTCLG
977275 .

\vspace{5mm}


\end{document}